\documentclass[journal]{IEEEtran}
\usepackage{cite}

\ifCLASSINFOpdf
  \usepackage[pdftex]{graphicx}
  \graphicspath{{/figures/}{../pdf/}{../jpeg/}}
  \DeclareGraphicsExtensions{.pdf,.jpeg,.png}
\else
   \usepackage[dvips]{graphicx}
  \graphicspath{{figures/}{../pdf/}{../jpeg/}}
  \DeclareGraphicsExtensions{.eps,.png}
\fi
\usepackage[cmex10]{amsmath}
\usepackage[]{algorithm}
\usepackage{algpseudocode}

\DeclareMathOperator*{\argmax}{\arg\!\max}
\DeclareMathOperator*{\argmin}{\arg\!\min}

\usepackage{amssymb}

\begin{document}

%
\title{ Direct Localization of Multiple  Sources by Partly Calibrated Arrays}
%
%
%

\author{~\IEEEmembership{}
        Amir Adler, \IEEEmembership{Member,~IEEE} and~Mati~Wax,~\IEEEmembership{Fellow,~IEEE}
\thanks{A. Adler  e-mail: adleram@mit.edu. M. Wax e-mail: matiwax@gmail.com.}
}

%
%

\markboth{SUBMITTED FOR PUBLICATION, IEEE}{}
%



\maketitle

\begin{abstract}
We present novel solutions to the problem of direct localization of multiple narrowband  and arbitrarily correlated  sources by partly calibrated arrays, i.e., arrays composed of  fully calibrated subarrays yet lacking  inter-array calibration.  The solutions presented vary in their performance and computational complexity. We present first a relaxed maximum likelihood solution whose  concentrated likelihood involves  \textit{only the unknown locations of the sources} and  requires an eigen-decomposition of the array covariance matrix at every potential location. To reduce the computational load, we introduce  an approximation which eliminates the need for such an   eigen-decomposition at every potential location. To further reduce the computational load, novel  MUSIC-like and MVDR-like solutions are presented which are computationally much simpler than   the existing solutions.     The performance of these solutions is evaluated and compared via simulations.

\end{abstract}


%
\IEEEpeerreviewmaketitle

\section{Introduction}
%
%
%
%
 \IEEEPARstart{P} artly  calibrated arrays are   arrays  composed of  \textit{fully calibrated subarrays, yet lacking inter-array calibration}. Such arrays are common in  large scale systems composed of small subarrays with large inter-array distances, as is the case in multi-site surveillance systems, multi-site communication systems, and multi-site radar systems. 
 
In such  arrays,  the large inter-array distances make the calibration of the whole array problematic.  Time and phase synchronization in such arrays may be problematic as well. If all the subarrays are time and phase synchronized, they are referred to as \textit{coherent}.  If each subarray is time and phase synchronized internally, but there is no time and phase synchronization  across the whole array, they are referred to as \textit {noncoherent}. If there is time synchronization across the whole array and  each subarray is time and phase synchronized internally,  but there are unknown phase offsets   between the subarrays, they are referred to as \textit{phase offset}.

A powerful  model for  partly calibrated arrays,  referred to as the Partly Calibrated Array (PCA) model, was introduced by  See and Gershman [] in the problem of direction finding. This model can cope with a variety of uncertainties in the direction finding problem, including  unknown subarrays displacements and unknown phase offsets between the subarrays.   It can be regarded as a generalization of a more limited model, introduced by Pesavento et. al [], addressing partly calibrated arrays composed of identically oriented subarrays with unknown subarray displacements. 

Apart from introducing the PCA model, [] presented a MUSIC-like technique for estimating the direction-of-arrival of multiple narrowband sources and the Cramer-Rao bound (CRB) for this problem. 
This work was followed by  Lie et al. [] and  Mavrychev et al. [] who introduced  MVDR-like techniques.  Liao and Chan [] exploited the special structure of the uniform linear array to reduce the computational complexity.  A sparse recovery approach for direction finding in partly calibrated arrays composed of subarrays with unknown displacements was introduced by Steffens and Pesavento []. 

Independently of this work on direction finding, Weiss [] and Weiss and Amar [] introduced  the PCA model in the direct localization problem, to cope with the unknown propagation  to the subarrays. Direct localization, advocated first in []-[] and further developed in []-[], is a localization scheme in which   the location is estimated directly from the data in one-step, as opposed to the more conventional two-step scheme, where  the directions-of-arrival to the subarrays are estimated in the first step and then, in the second step,  the location is estimated using triangulation. Direct localization provides not only higher accuracy at low signal-to-noise and low signal-to-interference ratios, but not less importantly, reduced ambiguity. This is because  the data association step,  needed in the two-step procedure  and  prone to ambiguity errors, is eliminated.    

Apart from introducing the PCA model,
[]  introduced the maximum likelihood solution for a single narrowband source, while [] extended this approach to widebanand sources,  introduced MUSIC-like solution for   sources with unknown waveforms and  a maximum likelihood solution for sources with known waveforms, as well as the CRB for these problems. This work was followed by Bosse et al. [], who introduced an alterantive  space-time approach for wideband sources, by  Delestre et al [] who extended this space-time approach to include the time delay information between the subarrays and by Tirer and Weiss [] who introduced MVDR-like technique for multiple wideband sources. There has been also extension to different types of signals and scattering. Reuven and Weiss []  to cyclostationary signals, Bar-Shalom and Weiss [] to environments with local scattering,  Yin et al [] to noncircular sources, and Yi et al. [] to signals parametrized by a small number of parameters. 

The PCA model has been used also in related localization problems. Closas et al. [] used it for direct localization in the  Global Navigation Satellites System (GNSS), to cope with the unknown propagation coefficients to the satellites, and derived the maximum likelihood estimator of the location.  Dmochowski et al. [] used it for direction finding of an acoustic source, to cope with uncertainties in the propagation environment and the acoustic array, and derived a modified version of the Steered  Minimum Variance  (STMV) [] method for a  wideband source. Bialkowski et al [], and subsequently Hack et al [], used it   in passive multistatic radar with widely separated receiving subarrays, to cope with  uncertainties   regarding the  propagation of the reference signal,  and  derived the maximum likelihood solution for a single moving target. Bar Shalom and Weiss [] used it for active radar involving multiple widely separated transmit antennas and multiple widely separated receiving arrays, to cope with uncertainties   regarding the propagation of the transmuted  signals, and derived maximum likelihood solutions  for locating a stationary target

In this paper we present novel solutions to the  problem of \textit{ direct localization} of narrowband  sources by partly calibrated arrays. We address the general case of arbitrarily correlated sources, including the case of \textit{fully correlated sources}, happening in coherent multipath propagation.  Note that since direction finding  can be considered as a special case of  direct localization, corresponding to the case that  the sources are in the far-field of the array, our solutions apply to both problems.

First, we present a  relaxed maximum likelihood solution which, by  eliminating all the nuisance parameters in the partly calibrated array model, reduces the problem to a concentrated likelihood involving \textit{only the $Q$ unknown locations} of the sources. The concentrated likelihood requires  an eigen-decomposition of the array covariance matrix for every potential location. In the special case of a single source with no multipath, this solution coincides with the  maximum likilihood solution []. Second, using the structure of the signal subspace, we  introduce an approximation  which eliminates  the need for an eigen-decomposition of the covariance matrix at every potential location, thus reducing significantly the computational load. In the special case of a single source with no multipath, this solution is computationally much simpler than the existing maximum likelihood solution [].  Third, to further reduce the computational complexity, we present MUSIC-like and MVDR-like solutions which, in contrast to  the existing MUSIC-like and MVDR-like solutions [], [], [], avoid the need for an eigen-decomposition for every potential location.

\indent The rest of the paper is organized as follows. The problem formulation is presented in section II.  Section III presents the "relaxed" maximum likelihood  solution, while section IV presents the reduced complexity solution and the MUSIC-like and MVDR-like solutions  which trade off performance for computational  load. The performance of the various solutions are compared using simulated data in section V. Finally, section VIII presents the  conclusions. 

\section{Problem Formulation}
Consider an array composed of $L$   fully calibrated subarrays, each composed  of $M_l$ antennas  with arbitrary locations and arbitrary directional characteristics. Let $M=\sum_{l=1}^L M_l $ denote the total number of antennas. Assume that $Q$ sources,  located at locations  $\{\mathbf p_q\}^{Q}_{q=1}$, with $\mathbf p_q\in R^{D\times 1}$, $D=1,2,3$, and emitting signals $\{s_q(t)\}^{Q}_{q=1}$, are impinging on the array.  

To capture both the direct localization and the direction finding problems, we allow the dimension $D$ to be a parameter. 
If the sources are in the far-field of the array then either $D=1$, if both the sources and the array are confined to a plane, or  $D=2$, if otherwise. In case the sources are in the near-field of the array, then either $D=2$, if both the sources and the array are confined to a plane, or $D=3$, if otherwise.

We further make the following assumptions regarding  the emitted signals, the array and the noise:

A1:  The number of sources $Q$ is \textit{known}.
 
A2: The emitted signals  are  \textit{narrowband}, i.e, their bandwidth is much smaller than the reciprocal  of the propagation time across the array, and  centered around angular frequency $\omega_c$.  

A3:  The emitted signals are  \textit{unknown}  with     zero mean    and arbitrary correlation, including being \textit{fully correlated},  as happens in \textit{coherent multipath} propagation. 

A4: The array is synchronized in time, but there is unknown phase offsets between the subarrays.

A5: The locations of the subarrays are  known, but with an  uncertainty of $\sigma_a^2 $.  

A6: The propagation model  is \textit{spherical waves} (it degenerates to plane waves if the sources are in the far-field of the array).

A7: The steering vectors of the subarrays toward any potential location $\mathbf p$, given by $\{\mathbf a_l(\mathbf p)\}^{L}_{l=1}$, are \textit{known} and have unit norm,  i.e., $\|\mathbf a_l(\mathbf p)\|=1$.

A8: The additive noises at the subarrays are  independent of the signals and independent of each other, and distributed as complex Gaussian with zero mean and covariance $\sigma_n^{2}\mathbf I_M $.

Assumptions A1-A3 and A6-A8 are  conventional and do not need further justification. Assumptions A4-A5 reflect  the current limitation of the Global Positioning System (GPS). A4 reflects the current accuracy of the GPS time data - typically 10 ns - which is good enough for time synchronization in the case  of narrowband signals, but not good enough    for phase synchronization.   A5 reflects the current accuracy of the GPS location data, which is typically 10 meters.

Under these assumptions, the PCA model for the $M_l\times 1$ vector of the complex envelopes of the received signals at the $l$-th subarray is given by
\begin{equation}
\mathbf x_{l}(t) = \sum_{q=1}^{Q} b_{l,q}\mathbf a_{l}(\mathbf p_q)  s_{q}(t- \tau_l(\mathbf p_q)) +\mathbf n_{l}(t),
\label{eq:snapshot}
\end{equation}
where $b_{l,q}$   is a complex coefficient associated with the propagation of the $q$-th signal to the $l$-th subarray,  $\mathbf a_l(\mathbf p_q)$ is the steering vector of the $l$-the subarray toward location $\mathbf p_q$, $j=\sqrt {-1}, $ $\tau_l(\mathbf p_q)$ is the delay from $\mathbf p_q$ to the $l$-th subarray,    and  $\mathbf n_{l}(t)$  is the noise at the $l$-th subarray.

The  partly calibrated nature of the array is embodied by the set of $QL$ complex coefficient $\{b_{l,q}\}$, $q=1,...,Q;\ l=1,...,L$,  assumed to be \textit{unknown} parameters. In our problem  these parameter capture the combined effect of the  unknown propagation  to the subarrays,  the  unknown subarrays displacement  due to subarrays location error,  and the  unknown phase offset between subarrays.  Though   $\{b_{l,q}\}$ are assumed here to be fixed in time, letting them vary over time can serve as a good model for the quasi-stationarity nature of some propagation channel, resulting from small temporal changes  due to movement of people, cars, trees, ect. [],[].

The narrowband assumption A2 implies that the time delays are well approximated by phase shifts, which allow us to rewrite (1) as  
\begin{equation}
\mathbf x_{l}(t) = \sum_{q=1}^{Q} b_{l,q}\mathbf a_{l}(\mathbf p_q)  s_{q}(t)e^{-j\omega_c \tau_l(\mathbf p_q)} +\mathbf n_{l}(t),
\label{eq:snapshot}
\end{equation}
Assuming the array is sampled $N$ times, we can express the received signals by the $l$-th subarray as 
\begin{equation}
\mathbf X_{l} =\mathbf A_{l}(\mathbf P_0)\mathbf B_l \mathbf S +\mathbf N_{l}, 
\label{eq:snapshot}
\end{equation} 
where $\mathbf X_l$ is the $M_l\times N$ matrix
\begin{equation}
\mathbf X_{l}= [\mathbf x_l(t_1), ..., \mathbf x_l (t_N)],
\label{eq:snapshot}
\end{equation}
$\mathbf A_l(\mathbf P_0)$ is the $M_l\times Q$ matrix of the steering vectors towards the $Q$ locations (to simplify the notation, the explicit dependence on the locations $\mathbf P_0=\{\mathbf p_1,...,\mathbf p_Q\}$ will be sometimes dropped)
\begin{equation}
\mathbf A_{l}(\mathbf P_0)=\mathbf A_l=[\mathbf a_l(\mathbf p_1)e^{-j\omega_c \tau_l(\mathbf p_1)}, ..., \mathbf a_l(\mathbf p_Q)e^{-j\omega_c \tau_l(\mathbf p_Q)}],
\label{eq:snapshot}
\end{equation}
$\mathbf B_l$ is a $Q\times Q$ diagonal matrix 
\begin{equation}
\mathbf B_{l} =diag(\mathbf b_l),
\label{eq:snapshot}
\end{equation}
with
\begin{equation}
\mathbf b_l=[b_{l,1},...,b_{l,Q}]^{T},
\label{eq:snapshot}
\end{equation}
$\mathbf S$ is the $Q\times N$ signals matrix
\begin{equation}
\mathbf S= [\mathbf s(t_1), ..., \mathbf s (t_N)]=\begin{pmatrix}\mathbf s_{1} \\
\vdots\ \\
\mathbf s_Q  \\
\end{pmatrix},
\label{eq:snapshot}
\end{equation}
with
\begin{equation}
\mathbf s(t)=[s_1(t),..., s_Q(t)]^{T},
\label{eq:snapshot}
\end{equation}
and $\mathbf N_l$ is the $M_l\times N$ matrix of the noise
\begin{equation}
\mathbf N_{l} =[\mathbf n_{l}(t_1),..., \mathbf n_{l}(t_N)],
\label{eq:snapshot}
\end{equation}

To equalize the contributions of the subarrays, we normalize their power, namely set
\begin{equation}
tr(\mathbf X_l \mathbf X_l^{H})=1,\;\;\;\;\ l=1,...,L
\label{eq:snapshot}
\end{equation}
where $ tr()$ denotes the trace operator and $H $ denotes the conjugate transpose.

\indent
We can now state the direct localization problem as follows: \textit{Given the received data $\{\mathbf X_l\}_{l=1}^{L}$, estimate the $Q$ locations $\{\mathbf p_q\}_{q=1}^{Q}$.}

\section{Relaxed Maximum Likelihood  Solution }

In this section we   derive the  Relaxed Maximum Likelihood (RML) solution.  

To this end, regarding the signals matrix $\mathbf S$ and the  coefficient matrices $\{\mathbf B_l\}$ as unknown   parameters, it follows from (3) and the Gaussian noise assumption A8 that the maximum likelihood cost function is  given by 

\begin{equation}
\label{eq:CM_criterion}
\mathbf  {\mathbf{\hat  P}} =\argmin_{\mathbf P, \{\mathbf B_l\},\mathbf S}   \sum_{l=1}^L \|\ \mathbf X_l-\mathbf A_l(\mathbf P) \mathbf B_l \mathbf S \|^{2}_{F}
\end{equation}
 Note that   this
 cost function is a multidimensional  nonlinear minimization  with a total of  $ DQ+2QL+2QN$ real unknown parameters, corresponding to $ \mathbf P, \{\mathbf B_l\}$, and $\mathbf S$, respectively. Out of this large number of unknowns, only the   $DQ $ unknowns corresponding to the locations  $\mathbf P$ are of our interest, while the other are considered as \textit{nuisance parameters}. 

As we show in Apendix A, the  exact solution of (12) yields a complicated expression which does not  seem to  enable the elimination of all the nuisance parameters. Consequently,  we next present a relaxed maximum likelihood solution which enables the  desired elimination and yields a concentrated likelihood involving only the unknown locations of the sources.  

Our first step is to eliminate  the unknown coefficients $\{\mathbf B_l\}$ by expressing them in terms of the other parameters $\mathbf P$ and $\mathbf S$. To this end, note that  $\mathbf B_l $ appears only in the $l$-th term in (12), implying that it can be estimated by the  following minimization problem:

\begin{equation}
\label{eq:R_i_test}
{\hat {\mathbf  B}_{l}} =\argmin_{\mathbf B_{l}} \|\ \mathbf X_l-\mathbf A_l  \mathbf B_l \mathbf S \|^{2}_{F}  \end{equation}
where we hold  $ \mathbf A_l$ and $\mathbf S$  fixed. Denoting by $J_l$  the cost function of (13), we have
\begin{eqnarray}
\label{eq:CM_criterion}
J_l & = & tr  (( \mathbf X_l-\mathbf A_l  \mathbf B_l \mathbf S)^{H}(\mathbf X_l-\mathbf A_l  \mathbf B_l \mathbf S))\nonumber\\
& = &  tr(\mathbf X^{H}_{l}\mathbf X_l)-tr(\mathbf B^{H}_{l}\mathbf A^{H}_{l}\mathbf X_l\mathbf S^{H})-tr(\mathbf A_l\mathbf B_l\mathbf S\mathbf X^{H}_{l}) \nonumber\\ & + &tr(\mathbf B^{H}_{l}\mathbf A^{H}_{l}\mathbf A_l\mathbf B_l\mathbf S\mathbf S^{H})
\end{eqnarray}
Dropping the terms which do not contain $\mathbf B_l$, we can rewrite it as
\begin{eqnarray}
J_l&=&-tr (\mathbf A_{l}\mathbf S\mathbf X^{H}_{l}diag(\mathbf b_l )) -tr(\mathbf A_l^H \mathbf X_l \mathbf S^H  \mathbf B_l^H)
\nonumber\\ &+& tr(\mathbf S\mathbf S^{H}\mathbf B^{H}_{l}\mathbf A^{H}_{l}\mathbf A_{l}{} diag(\mathbf b_l))  .
\end{eqnarray} 
Now,  equating  to zero the derivative with respect to $\mathbf b_l$, using  the  well known complex differentiation rules [] and the  following matrix differentiation rule [], 
\begin{equation}
 \frac{\partial}{\partial \mathbf b}tr( \mathbf A diag(\mathbf b))=diag (\mathbf A),
\end{equation}
 we get
\begin{equation}
\label{eq:R_i_test}
diag(\mathbf S\mathbf X^{H}_{l}\mathbf A_l) =  \mathbf diag(\mathbf S\mathbf S^{H}\mathbf B^{H}_{l}\mathbf A^{H}_l\mathbf A_l ).
\end{equation}
To solve this equation for $\mathbf B_l$, we first relax the equality of the diagonals of the two marices to an equality of the whole matrices, yielding  
\begin{equation}
\label{eq:R_i_test}
\mathbf S\mathbf X^{H}_{l}\mathbf A_l =  \mathbf S \mathbf S^{H}{\mathbf{  B}^{H}_{l}\mathbf A^{H}_l\mathbf A_l }.
\end{equation}
Next, we relax the constraint that $\mathbf B_l$  is  diagonal and allow it to be an arbitrary matrix, which enables us  to  straightfowardly solve this equation for $\mathbf B_l$, yielding 
\begin{equation}
\label{eq:R_i_test}
{\mathbf{\hat  B}_{l}} =  \mathbf (\mathbf A^{H}_l\mathbf A_l )^{-1}\mathbf A_l^{H}  \mathbf X_l \mathbf S^{H}(\mathbf S\mathbf S^{H})^{-1}.
\end{equation}
Now, multiplying  from the left and right by $\mathbf A_l$ and $\mathbf S$, respectively, we get
\begin{equation}
\label{eq:R_i_test}
 \mathbf A_l {\mathbf{ \hat B}_l} \mathbf S =  \mathbf A_l(\mathbf A^{H}_l\mathbf A_l )^{-1}\mathbf A_l^{H}  \mathbf X_l \mathbf S^{H}(\mathbf S\mathbf S^{H})^{-1} \mathbf S
\end{equation}
or alternatively, 
\begin{equation}
\label{eq:R_i_test}
 \mathbf A_l \mathbf {\hat  B}_l \mathbf S =  \mathbf{ P_{\mathbf A_l}}\mathbf X_l \mathbf  P_{\mathbf S^{H}}
\end{equation}
where $\mathbf{P_{\mathbf A_l}} $ is the projection matrix on column span of of $\mathbf A_l $
\begin{equation}
\label{eq:R_i_test}
\mathbf{  P_{\mathbf A_l}}= \mathbf A_l(\mathbf A^{H}_l\mathbf A_l )^{-1}\mathbf A_l^{H}  \mathbf , 
\end{equation}
and $\mathbf P_{\mathbf S^{H}}$ is the projection matrix on the column span of $\mathbf S^{H} $
\begin{equation}
\mathbf P_{\mathbf S^{H}}=\mathbf S^{H}(\mathbf S\mathbf S^{H})^{-1} \mathbf S.
\end{equation}
Substituting (21) into (12), yields\begin{equation}
\label{eq:R_i_test}
{\mathbf{\hat  P}} =\argmin_{\mathbf P, \mathbf S}  \sum_{l=1}^{L}\|\ \mathbf X_l-\mathbf {P}_{\mathbf A_l(\mathbf P)} \mathbf X_{l} \mathbf  P_{\mathbf S^{H}}\|^{2}_{F},
\end{equation}
which, using the properties of the trace operator and  the  projection matrix,  with some straightforward manipulations, reduces to    
\begin{equation}
\label{eq:R}
\mathbf  {\mathbf{\hat P}} =\argmax_{\mathbf P, \mathbf S}tr(\mathbf P_{\mathbf S^{H}}  \sum_{l=1}^{L}\mathbf X_l^{H} \mathbf{P}_{\mathbf A_l(\mathbf P)} \mathbf X_l)
\end{equation}
This expression can be interpreted as a search for the locations $\mathbf P$ and the signal matrix $\mathbf S$  for which there is maximum correlation between the signal subspace defined by $\mathbf P_{S^{H}}$ and the  sum of projections of $\mathbf X_l$ on the signal subspaces defined by $\mathbf P_{\mathbf A_l(\mathbf P)}$, $l=1, ... ,L$.

To further eliminate the unknowns parameters of  the matrix $\mathbf S$, we next evaluate (25) separately  for noncoherent and coherent signals. 
\subsection { Noncoherent Signals}

 In case the signals are  noncoherent,  the signal subspace defined by $\mathbf P_{\mathbf S^{H}}$ is $Q$-dimensional. This, in turn, implies that we can express  $\mathbf P_{\mathbf S^{H}}$  as  

\begin{equation}
\label{eq:R_i_test}
\mathbf P_{\mathbf S^{H}}=  {\tilde {\mathbf S}}^{H} {\tilde{\mathbf S}} , 
\end{equation}
where ${\tilde{\mathbf S}}$ obeys 
\begin{equation}
\label{eq:W} 
{\tilde{\mathbf S}}{\tilde{\mathbf S}}^{H}= \mathbf I_Q
\end{equation}
where $\mathbf I_Q$ is the $Q\times Q$ identity matrix.  Substituting this expression in (25), using the properties of the trace operator, we get  

\begin{equation}
\label{eq:W} 
\mathbf  {\mathbf{\hat  P}} =\argmax_{\mathbf P, \tilde{ \mathbf S}; \; \tilde{ \mathbf S}\tilde{ \mathbf S}^{H} =\mathbf I_q}tr({{ \tilde {\mathbf S}}}(  \sum_{l=1}^{L}\mathbf X_l^{H} \mathbf{  P}_{\mathbf A_l(\mathbf P)}\mathbf X_l{ \tilde {\mathbf S}^{H})}  
\end{equation}
Maximizing this expression    over ${ \tilde {\mathbf S}}$, holding $\mathbf P$ constant,  we get 
\begin{equation}
\label{eq:R_i_test}
\hat {\tilde {\mathbf  S}}^H(\mathbf P)=\tilde{\mathbf V}_S(\mathbf P)=[\tilde {\mathbf v}_1,...,\tilde {\mathbf v}_Q ] , 
\end{equation}   
where $\tilde {\mathbf  { v}}_q$ denotes the $N\times 1$ eigenvector corresponding to the  $q$-th eigenvalue of the $N\times N$  matrix $  \sum_{l=1}^{L}\mathbf X_l^{H} \mathbf{  P}_{\mathbf A_l(\mathbf P)}\mathbf X_l)$. Substituting this expression for $\hat {\tilde {\mathbf  S}}(\mathbf P)$ back into (28), we get       
\begin{equation}
\label{eq:W} 
\mathbf  {\mathbf{\hat  P}} =\argmax_{\mathbf P }\sum_{q=1}^{Q}\lambda_q ( \sum_{l=1}^{L} {\mathbf X_l^{H}} \mathbf{  P}_{\mathbf A_l(\mathbf P)}\mathbf X_l) 
\end{equation}
where $\lambda_q()$ denotes the $q$-th eigenvalue of the bracketed matrix. 

For large $N$, computing the eigenvalues of the  $N\times N$ matrix $\mathbf X_l^{H} \mathbf{  P_{\mathbf A_l(\mathbf P)}}\mathbf X_l$ may be prohibitive. We next show how to reduce the dimensionality of this problem. 

To this end,  
denote by  $ \mathbf P_{\mathbf A(\mathbf P)}$  the $M\times M$ block-diagonal matrix
\begin{equation}
\mathbf P_{\mathbf A(\mathbf P)} = \begin{pmatrix}\mathbf P_{\mathbf A_{1}(\mathbf P)} & \cdots & \mathbf 0 \\
\vdots & \ddots & \vdots \\
\mathbf 0 & \cdots  & \mathbf P_{\mathbf A_{L}(\mathbf P)} \\
\end{pmatrix},
\label{eq:snapshot}
\end{equation}
by $\mathbf X$ is the $M\times N $ matrix of the sampled data
\begin{equation}
\mathbf X= (\mathbf X_1^T, \dots, \mathbf X_L^T)^T,
\end{equation}
and by $\hat{\mathbf R}=\mathbf X\mathbf X^H$  the $M\times M$ sample-covariance  matrix of the array
\begin{equation}
\hat{\mathbf R} = \begin{pmatrix}\mathbf X_1\mathbf X_1^{H} & \cdots & \mathbf X_1\mathbf X_L^{H} \\
\vdots & \ddots& \vdots \\
\mathbf X_1\mathbf X_L^{H} & \cdots & \mathbf X_L\mathbf X_L^{H} \\
\end{pmatrix}= \begin{pmatrix}\hat{\mathbf R}_{1,1} & \cdots & \hat{\mathbf R}_{1,L}\\
\vdots & \ddots& \vdots \\
\hat{\mathbf R}_{L,1}& \cdots & \hat{\mathbf R}_{L,L}\\
\end{pmatrix}.
\label{eq:snapshot}
\end{equation}
Now, as we show in Appendix B,
\begin{equation}
\label{eq:W} 
\lambda_q(\sum_{l=1}^{L} {\mathbf X_l^{H}} \mathbf{  P}_{\mathbf A_l(\mathbf P)}\mathbf X_l )=\lambda_q ( \mathbf P_{\mathbf A(\mathbf P)} \hat{\mathbf R} \mathbf P_{\mathbf A(\mathbf P)}),
 \end{equation}
which when substituted into  (30) yields 
\begin{equation}
\label{eq:W} 
\mathbf {\mathbf{\hat P}} =\argmax_{\mathbf p} \sum_{q=1}^{Q} \lambda_q ( \mathbf P_{\mathbf A(\mathbf P)} \hat{\mathbf R} \mathbf P_{\mathbf A(\mathbf P)}).
 \end{equation}
To further simplify this expression, let $\tilde{\mathbf A}$ denote the block-diagonal matrix
\begin{equation}
\tilde {\mathbf A} = \begin{pmatrix} \tilde { \mathbf A}_1 & \cdots & \mathbf 0 \\
\vdots & \ddots & \vdots \\
\mathbf 0 & \cdots  & \tilde {\mathbf A}_L \\
\end{pmatrix},
\label{eq:snapshot}
\end{equation}
where $\tilde {\mathbf A}_l$ is given by
\begin{equation}
\label{eq:R_i_test}
\tilde {\mathbf A}_l=  {\mathbf A}_l( {\mathbf A}_l^{H} {\mathbf A}_l)^{-1/2}. 
\end{equation}
Using this notation  we can rewrite $\mathbf P_{\mathbf A}$ as
\begin{equation}
\mathbf P_{\mathbf A} =\tilde {\mathbf A}\tilde {\mathbf A}^{H}=\begin{pmatrix}\tilde {\mathbf A}_1\tilde {\mathbf A}^{H}_{1} & \cdots & \mathbf 0 \\
\vdots & \ddots & \vdots \\
\mathbf 0 & \cdots  & \tilde {\mathbf A}_L\tilde {\mathbf A}^{H}_{L}  \\
\end{pmatrix},
\label{eq:snapshot}
\end{equation}
which implies, using  the invariance of the eigenvalues of a product of matrices to their cyclic permutation [], that
\begin{equation}
\label{eq:W} 
\lambda_q ( \mathbf{  P}_{\mathbf A} \hat{\mathbf R} \mathbf{  P_{\mathbf A}})=\lambda_q ( \mathbf{  P_{\mathbf A}} \mathbf{  P_{\mathbf A}} \hat{\mathbf R})=\lambda_q ( {  \mathbf{  P_{\mathbf A}}} \hat{\mathbf R}) =\lambda_q ( \tilde {\mathbf A}^{H} \hat{\mathbf R}\tilde {\mathbf A})  
\end{equation}
Substituting this result  into (35), we get 
\begin{equation}
\label{eq:W} 
\mathbf  {\mathbf{\hat  P}} =\argmax_{\mathbf P }\sum_{q=1}^{Q}\lambda_q ( \tilde {\mathbf A}^{H}(\mathbf P) \hat{\mathbf R} \tilde {\mathbf A}(\mathbf P))
 \end{equation}
Note that since the matrix   
$\tilde {\mathbf A}^{H}(\mathbf P)\hat{\mathbf R} \tilde {\mathbf A}(\mathbf P)$ is    $LQ\times LQ$, and since typically $LQ\ll N$,  the computational complexity of the solution (40) is significantly smaller than that of (30).
Yet,  the complexity of this solution is still high, as it  involves the computation of the $Q$ largest eigenvalues of this matrix for every potential location  $\mathbf P$, and a $Q$-dimensional search for the location $\mathbf P $ for which this  sum of eigenvalues is maximized.

To reduce the computational load of the $Q$-dimensional search over $\mathbf P $, we can employ the Alternative Projection (AP) algorithm [], which  transforms a Q-dimensional search into an iterative process involving only \textit{single source} searches. 

Denote the cost function by 
\begin{equation}
\label{eq:W} 
g({\mathbf P)}=\sum_{q=1}^{Q}\lambda_q ( \tilde {\mathbf A}^{H}(\mathbf P)\hat{\mathbf R} \tilde {\mathbf A}(\mathbf P))
 \end{equation}
The AP algorithm involves two phases. In the first phase, referred to as initialization, the number of sources is increased from $q=1$  to $q=Q$, with  the $q$-th step involving a  maximization  over $\mathbf p_q$, with the other $q-1$ pre-computed locations held fixed:
\begin{equation}
\label{eq:W} 
\hat {\mathbf p}_q =\argmax_{\mathbf P_q} g(\mathbf P_q^{(0)}) 
 \end{equation}
 where
\begin{equation}
\label{eq:R_i_test}
\mathbf P_q^{(0)}=( \hat {\mathbf p}_1,...,\hat {\mathbf p}_{q-1},\mathbf p_q) , 
\end{equation} 
In the second phase,  the algorithm involves multiple iterations till convergence, with the $k+1$ iteration for the $q$-th source given by
\begin{equation}
\label{eq:W} 
{{\hat {\mathbf  p}^{(k+1)}_{q}}} =\argmax_{\mathbf p_q }g( \mathbf P^{(k)}_{q})
 \end{equation}
 where
\begin{equation}
\label{eq:R_i_test}
\tilde {\mathbf P}^{(k)}_{q}=( \hat {\mathbf p}^{(k)}_{1},...,\hat {\mathbf  p}_{q-1}^{(k)},\mathbf p_q, \hat {\mathbf p}_{q+1}^{(k)},...,\hat {\mathbf p}_Q^{(k)}). 
\end{equation}

\subsection { Coherent Signals} 

When the signals are coherent,  the raws of the signals matrix  $\mathbf S$ are identical hence  we have,
\begin{equation}
\mathbf s_1=\mathbf s_2= ...=\mathbf s_Q =\mathbf s. 
\end{equation}
It then follows  that  $\mathbf P_{\mathbf S^{H}}$, the projection matrix on the signal subspace, is in this case rank-1 and given by
\begin{equation}
\mathbf P_{\mathbf S^{H}}=\mathbf {s}^{H}(\mathbf {s}\mathbf {s}^{H})^{-1} \mathbf {s}=\tilde{\mathbf s}^{H}\tilde{\mathbf s},
\end{equation}
which when substituted   into (25), yields
\begin{equation}
\label{eq:W} 
\mathbf{\mathbf{\hat P}} =\argmax_{\mathbf{P},\tilde{\mathbf s}; \;  \tilde{\mathbf s} \tilde{\mathbf s}^{H}=1}
tr( \tilde{\mathbf s}(  \sum_{l=1}^{L}\mathbf X_l^{H} \mathbf{  P}_{\mathbf A_l(\mathbf P)}\mathbf X_l){ \tilde{\mathbf s}^{H}).}  
\end{equation}
Maximizing this expression over $\tilde{\mathbf s}$, while holding $\mathbf P$ constant,  yields 
\begin{equation}
\label{eq:R_i_test}
\hat{\tilde{\mathbf  s}}^H(\mathbf P)=\tilde {\mathbf v}_1, 
\end{equation}   
Substituting this result back into (48) we get       
\begin{equation}
\label{eq:W} 
\mathbf  {\mathbf{\hat  P}} =\argmax_{\mathbf P }\lambda_1 ( \sum_{l=1}^{L} {\mathbf X_l^{H}} \mathbf{  P}_{\mathbf A_l(\mathbf P)}\mathbf X_l).
\end{equation}
Using (34), this becomes
 \begin{equation}
\label{eq:W} 
\mathbf  {\mathbf{\hat  P}} =\argmax_{\mathbf P }\lambda_1 (\mathbf P_{\mathbf A(\mathbf P)}\hat {\mathbf R }\mathbf P_{\mathbf A(\mathbf P}), 
\end{equation}
which can be rewritten as
\begin{equation}
\label{eq:W} 
\mathbf  {\mathbf{\hat  P}}  = \argmax_{\mathbf P }\lambda_1 (\tilde {\mathbf A}^{H}(\mathbf P) \hat{\mathbf R} \tilde {\mathbf A}(\mathbf P)).
\end{equation}

This expression is similar to that obtained for the noncoherent signals (35), with the difference that here only the first eigenvalue of ${\mathbf A}^{H}(\mathbf P) \hat{\mathbf R} \tilde {\mathbf A}(\mathbf P)$ is involved. The complexity of this solution is still high, as it  involves the computation of the  largest eigenvalue of this matrix for every potential locations  $\mathbf P$, and a $Q$-dimensional search over $\mathbf P $ for the $Q$ potential locations for which this  sum of eigenvalues is maximized.
To simplify the computational load of the $Q$-dimensional search over $\mathbf P $,  one can use the  AP algorithm  described in (43)-(47), with the difference  being that $g(\mathbf P)=\lambda_1 ( \tilde {\mathbf A}^{H}(\mathbf P) \hat{\mathbf R} \tilde {\mathbf A}(\mathbf P))$.  

The solution (52) admits a beamforming interpretation. 
To reveal it,  first note that by the definition of the largest eigenvector, we can rewrite it as 
\begin{equation}
\label{eq:W} 
\hat {\mathbf p} =\argmax_{ \mathbf P,\mathbf w ;\;\mathbf w^{H} \mathbf w=1 }\mathbf w^{H} \tilde {\mathbf A}^{H}(\mathbf P)\hat{\mathbf R} \tilde {\mathbf A} (\mathbf P)\mathbf w 
\end{equation}
where $\mathbf w$ is the $LQ\times 1$ vector
\begin{equation}
\mathbf w=[\mathbf w_1^T,...,\mathbf w_L^T]^{T}.
\label{eq:snapshot}
\end{equation}
with
\begin{equation}
\mathbf w_l=[w_1,...,w_Q]^{T}.
\label{eq:snapshot}
\end{equation}
Now, it can be readily verified that
\begin{eqnarray}
\label{eq:Q} 
\mathbf w^{H}\tilde {\mathbf A}^{H}\hat{\mathbf R }\tilde {\mathbf A}\mathbf w & = & \sum_{l,k=1}^{L}  \mathbf w^{H}_{l} 
\tilde {\mathbf A}_l^{H}\hat{\mathbf R}_{l,k}\tilde {\mathbf A}_{k} \mathbf  w_{k}\cr 
 &=& \sum_{n=1}^{N}|\sum_{l=1}^{L} \mathbf w^{H}_{l}  \mathbf  {\tilde {\mathbf A}}_l^{H}  \mathbf x_{l}(t_n)|^{2},\;\;\;\;
\end{eqnarray} 
which implies that
\begin{equation}
\hat {\mathbf  P} =\argmax_{ \mathbf P,\mathbf w ;\;\mathbf w^{H} \mathbf w=1 }\sum_{n=1}^{N}|\sum_{l=1}^{L} \mathbf w^{H}_{l}  \mathbf  {\tilde {\mathbf A}}_l^{H}(\mathbf P) \mathbf x_{l}(t_n)|^{2}.
\end{equation}
Note that $\mathbf {\tilde {\mathbf A}}_l^{H}(\mathbf P) \mathbf x_{l}(t_n)$ can be interpreted as beamforming at  the $l$-th subarray towards locations $\mathbf P$, while $\sum^{L}_{l} \mathbf w_l^{H}\mathbf {\tilde {\mathbf A}}_l^{H}(\mathbf P) \mathbf x_{l}(t_n)$ can be interpreted as a second level of beamforming,  aimed at combing coherently the outputs of the  subarrays' beamformers.  The whole expression can therefore be interpreted as a search for the weights $\mathbf w$ and locations $\mathbf P$ for which the power output of this  two-level beamforming is maximized.

\subsection { Single Signal}

In the case of a single signal, the matrix $\mathbf A$ reduces to

\begin{equation}
\mathbf A(\mathbf p) = \tilde{\mathbf A}(\mathbf p)=\begin{pmatrix}  \tilde{\mathbf a}_1(\mathbf p)& \cdots & \mathbf 0 \\
\vdots & \ddots & \vdots \\
\mathbf 0 & \cdots  & \tilde{\mathbf a}_L(\mathbf p)\\
\end{pmatrix},
\label{eq:snapshot}
\end{equation}
where
\begin{equation}
\label{eq:W} 
\tilde{\mathbf a}_l(\mathbf p) =\mathbf a_l(\mathbf p)e^{-j\omega_c \tau_l(\mathbf p)},
\end{equation}
and (52) becomes
\begin{equation}
\label{eq:W} 
 \hat {\mathbf  p} =\argmax_{ \mathbf p }\lambda_1 (\tilde {\mathbf A}^{H}(\mathbf p)  \hat{\mathbf R} \tilde {\mathbf A} (\mathbf p)),
\end{equation}
which is identical to the expression derived by Weiss []. 

As in the coherent signals case, this expression has  a beamforming  interpretation. Indeed, following the same steps leading from (53) to (57), we get
\begin{equation}
\hat {\mathbf  p} =\argmax_{ \mathbf p,\mathbf w ;\;\mathbf w^{H} \mathbf w=1 }\sum_{n=1}^{N}|\sum_{l=1}^{L}  w^{H}_{l}  \mathbf  {\tilde {\mathbf a}}_l^{H}(\mathbf p)  \mathbf x_{l}(t_n)|^{2}
\end{equation}
Note that $\mathbf {\tilde {\mathbf a}}_l^{H} \mathbf x_{l}(t_n)$ can be interpreted as beamforming at  the $l$-th subarray towards location $\mathbf p$, while $\sum^{L}_{l} w_l^{H}\mathbf {\tilde {\mathbf a}}_l^{H}(\mathbf p) \mathbf x_{l}(t_n)$ can be interpreted as a second level of beamforming,  aimed at combining coherently the outputs of the  subarrays' beamformers.  The whole expression can therefore be interpreted as a search for the weights $\mathbf w$ and location $\mathbf p$ for which the power output of this  two-level beamforming is maximized.

\section{Reduced Complexity Solutions }

The RML solution derived above is computationally complex.
In this section we present two reduced complexity solutions with different level of complexity.

\subsection {  Reduced Complexity Signal Subspace Solution } 

We first  present a solution which simplifies considerably the computation but still requires a $Q$-dimensional maximization, based on exploiting the structure of the signal subspace.  

To reveal the structure of the signal subspace,  let us rewrite $\mathbf X $ as
\begin{equation}
\mathbf X = \mathbf Y +\mathbf N, 
\label{eq:snapshot}
\end{equation} 
where $\mathbf Y$ is the signal component and $\mathbf N$ is the noise. From (3), we can express $\mathbf Y$ as

\begin{equation}
\mathbf Y =
\begin{pmatrix} \mathbf Y_1  \\
\vdots  \\
\mathbf Y_L  \\
\end{pmatrix}=
\begin{pmatrix} \mathbf A_1  \mathbf B_1\mathbf S\\
\vdots  \\
\mathbf A_L\mathbf B_L\mathbf S  \\
\end{pmatrix}=\mathbf A\overline{\mathbf S}
\label{eq:snapshot}
\end{equation}
where $\mathbf A$ is the $M\times QL $ block-diagonal matrix
\begin{equation}
\mathbf A 
=\begin{pmatrix}  \mathbf A_1 & \cdots & \mathbf 0 \\
\vdots & \ddots & \vdots \\
\mathbf 0 & \cdots  & \mathbf A_L \\
\end{pmatrix},
\label{eq:snapshot}
\end{equation}
and $\overline{\mathbf S}$  is the $ QL\times N$ matrix given by
\begin{equation}
\overline{\mathbf S }=
\begin{pmatrix}   \mathbf B_1\mathbf S\\
\vdots  \\
\mathbf B_L\mathbf S  \\
\end{pmatrix}.
\label{eq:snapshot}
\end{equation}
Substituting (63) into (62) we get
\begin{equation}
\mathbf X=\mathbf A\overline{\mathbf S}+\mathbf N.
\end{equation}
Now, denoting by $\mathbf R$ and  $\mathbf R_{\overline {\mathbf S}}$ the covariance matrices of $\mathbf X$ and $\overline{\mathbf S}$, respectively, it follows from (66) and from assumptions A4 and A8 regarding the  properties of the signals and noise that
\begin{equation}
\mathbf R=\mathbf A \mathbf R_{\overline{\mathbf S}}\mathbf A^H+\sigma^2\mathbf I_M.
\end{equation}
Multiplying this equation from both sides by  $\mathbf P_{\mathbf A}$, we get
\begin{equation}
\mathbf P_{\mathbf A}\mathbf R\mathbf P_{\mathbf A}=\mathbf P_{\mathbf A}\mathbf A\mathbf R_{\overline{\mathbf S}} \mathbf A^H\mathbf P_{\mathbf A} +\sigma^2\mathbf P_{\mathbf A}\mathbf I_M\mathbf P_{\mathbf A}.
\label{eq:snapshot}
\end{equation} 
Recalling that 
\begin{equation}
\mathbf P_{\mathbf A}\mathbf A=\mathbf A,
\end {equation}
this becomes
\begin{equation}
\mathbf P_{\mathbf A}\mathbf R\mathbf P_{\mathbf A}=\mathbf A\mathbf R_{\overline{\mathbf S}}\mathbf A^H +\sigma^2\mathbf P_{\mathbf A}, 
\label{eq:snapshot}
\end{equation} 
Comparing (67) and (70), it is clear  that both  $\mathbf R$ and $\mathbf P_{\mathbf A}\mathbf R \mathbf P_{\mathbf A} $ have the same $Q$-dimensional signal subspace $\mathbf A\mathbf R_{\overline{\mathbf S}}\mathbf A^H$. Since this subspace is spanned by the $Q$ largest eigenvectors of $\mathbf R$, we have  
\begin{equation}
\label{eq:R_i_test}
\mathbf {v}_q(\mathbf P_{\mathbf A}  \mathbf R  \mathbf P_{\mathbf A} \mathbf)=\mathbf v_q (\mathbf R)=\mathbf v_q,\;\;\;\;q=1,...,Q. 
\end{equation}
or more explicitly, inserting the  explicit dependence on the \textit{true location} $\mathbf P_0$, we have
\begin{equation}
\label{eq:R_i_test}
\mathbf {v}_q(\mathbf P_{\mathbf A(\mathbf P_0)}  \mathbf R  \mathbf P_{\mathbf A(\mathbf P_0)} \mathbf)=\mathbf v_q (\mathbf R)=\mathbf v_q,\;\;\;\;q=1,...,Q. 
\end{equation}
We will next exploit this relation to simplify the computational load of the RML solution. 

To this end, from (35), using
 the well known properties of the eigenvectors, we can write
\begin{equation}
\label{eq:W} 
\mathbf  {\mathbf{\hat  P}} =\argmax_{\mathbf P  }
tr(\tilde { \mathbf V}_S^{H}(\mathbf P) \mathbf P_{\mathbf A(\mathbf P)} \hat{\mathbf R}  \mathbf P_{\mathbf A(\mathbf P)}\tilde {\mathbf V}_S(\mathbf P)).
\end{equation}
where $\tilde {\mathbf V}_S(\mathbf P)$ is the $M\times Q $ matrix of the $Q$ largest eigenvectors of the matrix $\mathbf P_{\mathbf A(\mathbf P)} \hat{\mathbf R  } \mathbf P_{\mathbf A(\mathbf P)}$.

Now,  since $\hat{\mathbf P} $,  the maximizing value of (73), is  \textit{close to true value} $\mathbf P_0$, and since $\hat{\mathbf R}$ is close to $\mathbf R$ - the error in these two approximations diminishes as the number of samples grows - it follows from (72) that
\begin{equation}
\label{eq:R_i_test}
\mathbf {v}_q(\mathbf P_{\mathbf A(\hat{\mathbf P})} \hat{ \mathbf R}  \mathbf P_{\mathbf A(\hat{\mathbf P})} \mathbf)\approx\mathbf v_q (\hat{\mathbf R})=\hat{\mathbf {v}}_q\;\;\;\;q=1,...,Q, 
\end{equation}
where $\hat {\mathbf  { v}}_q$ denotes the  eigenvector corresponding to the  $q$-th eigenvalue of the   matrix $\hat{\mathbf R}$. This implies that 
\begin{equation}
\mathbf V_S(\hat{\mathbf P})ֿ\approx\hat{\mathbf V}_S,
\end{equation}
where $\hat{\mathbf V}_S$ is the matrix of the $Q$ largest eigenvectors of $\hat {\mathbf R}$,
\begin{equation}
\label{eq:R_i_test}
  \hat{\mathbf V}_{S}=[\hat{{\mathbf v}}_1,...,\hat{\mathbf v}_{Q} ]. 
\end{equation}
This in turn implies  that the maximum  value of (73) is unchanged if $\tilde{\mathbf V}_S(\mathbf P)$ is replaced by $\hat{\mathbf V}_S$, implying that the maximization problem (73) can be reformulated as
\begin{equation}
\label{eq:W} 
\hat{\mathbf P}=\argmax_{\mathbf P}tr(\hat{\mathbf V}_S^{H} {\mathbf P_{\mathbf A(\mathbf P	)} }\hat{\mathbf  R}  \mathbf P_{\mathbf A(\mathbf P)} \hat{\mathbf V}_S).
 \end{equation}
In the coherent signals case, this becomes
\begin{equation}
\label{eq:W} 
\mathbf  {\mathbf{\hat  P}} =\argmax_{\mathbf P }
tr( \hat{\mathbf v}_1^{H} \mathbf P_{\mathbf A(\mathbf P)} \hat{\mathbf R}  \mathbf P_{\mathbf A(\mathbf P)}\hat{\mathbf v}_1).
\end{equation}
and in the single signal case, this reduces to
\begin{equation}
\label{eq:W} 
\mathbf  {\mathbf{\hat  p}} =\argmax_{\mathbf p  }
tr(\hat{\mathbf v}_1^{H} \mathbf P_{\mathbf A (\mathbf p)}\hat{\mathbf R}  \mathbf P_{\mathbf A(\mathbf p)}\hat{\mathbf v}_1).
\end{equation}

Note that the computational complexity of (77), (78) and (79) is significantly lower than those of the corresponding  solutions  (35), (53) and (60), since here \textit{ no eigen-decomposition} is required  for every 
potential location. This implies also a significant computational saving  over the existing solution for the  case of \textit{a single signal} presented by Weiss [].

The expression (78) for coherent signals admits  a beamforming interpretation.  To see it, let  $\hat{\mathbf v}_1$ be segmented into its $L$ subvectors corresponding to the $L$ subarrays
\begin{equation}
\label{eq:R_i_test}
\hat{\mathbf v}_{1}=[\hat{\mathbf v}^{T}_{1_1},...,\hat{\mathbf v}^{T}_{1_L}]^{T},  
\end{equation}
where $\hat{\mathbf {v}}_{1_l}$is is the $l$-th segment of $\hat{\mathbf v}_1$. Substituting it into (74) with some straightforward manipulation,  yield
\begin{equation}
\label{eq:W} 
\mathbf{\hat  P} =\argmax_{\mathbf P} \sum_{n=1}^{N}|\sum_{l=1}^{L}  (\mathbf{  P}_{\mathbf A_l(\mathbf P)}\hat{\mathbf v}_{1_l})^{H}  \mathbf x_{l}(t_n)|^{2}, 
\end{equation}
This expression can be  interpreted as a two-step beamforming. In the first step,  a set of beamformers are applied at the subarrays, which are based on the largest eigenvector $\hat{\mathbf v}_1$ and given by $\mathbf w_{l}=\mathbf P_{\mathbf A_{l}(\mathbf P)}\hat{\mathbf v}_{1_l}$. Then in the second step, the total beamformer power output is maximized over all potential location. 
As we show in Appendix C, the beamformer $\mathbf w_{l}=\mathbf P_{\mathbf A_{l}(\mathbf P)}\hat{\mathbf v}_{1_l}$ compensates, in a suboptimal way, for the unknown $\mathbf {b}_l$.  

Expressions (77)-(79) can be further simplified. To this end, we first rewrite (77), using the properties of the trace operator, as  
\begin{equation}
\label{eq:W} 
\hat{\mathbf P}=\argmax_{\mathbf P}tr(\hat{\mathbf V}_S\hat{\mathbf V}_S^{H} {\mathbf P_{\mathbf A(\mathbf P	)} }\hat{\mathbf  R} \mathbf P_{\mathbf A(\mathbf P)}) .
 \end{equation}
Now,  from the eigen-decomposition of $\hat{\mathbf R} $ we have
\begin{equation}
\hat{\mathbf R}=\sum_{m=1}^M \hat{\lambda}_m\hat{\mathbf v}_m\hat{\mathbf v}_m^H,
\end{equation}
where $\hat{\lambda}_m$ denotes $m$-th eigenvalue of $\hat{\mathbf R}$,  and similarly

\begin{equation}
\hat{\mathbf V}_S\hat{\mathbf V}_S^H =\sum_{q=1}^Q \hat{\mathbf v}_q\hat{\mathbf v}_q^H.
\end{equation}
Substituting these expression into (82), with some straightforward manipulations, we get
\begin{equation}
\label{eq:W} 
\hat{\mathbf P}=\argmax_{\mathbf P} \sum_{m=1}^M\sum_{q=1}^Q \hat{\lambda}_m|\hat{\mathbf v}_m^H \mathbf P_{\mathbf A(\mathbf P)}\hat{\mathbf v}_q|^2,
\end{equation}
which can be interpreted as a weighted projection of the eigenvectors on the signal subspace defined by $\mathbf P_{\mathbf A(\mathbf P)}$, with the weights given by the corresponding eigenvalues.

Using (31), (85)  can be further simplified to
\begin{equation}
\label{eq:W} 
\hat{\mathbf P}=\argmax_{\mathbf P} \sum_{l=1}^L\sum_{m=1}^M\sum_{q=1}^Q \hat{\lambda}_m|\hat{\mathbf v}_{m_l}^H \mathbf P_{\mathbf A_l(\mathbf P)}\hat{\mathbf v}_{q_l}|^2.
\end{equation}
where $\hat{\mathbf {v}}_{m_l}$ is  the $l$-th segment of $\hat{\mathbf v}_m$. 

For the case of coherent sources this reduces to
\begin{equation}
\label{eq:W} 
\hat{\mathbf P}=\argmax_{\mathbf P} \sum_{l=1}^L\sum_{m=1}^M\hat{\lambda}_m|\hat{\mathbf v}_{m_l}^H \mathbf P_{\mathbf A_l(\mathbf P)}\hat{\mathbf v}_{1_l}|^2,
\end{equation}
and for a single signal this becomes
\begin{equation}
\label{eq:W} 
\hat{\mathbf p}=\argmax_{\mathbf p} \sum_{l=1}^L\sum_{m=1}^M \hat{\lambda}_m|\hat{\mathbf v}_{m_l}^H \mathbf P_{\mathbf a_l(\mathbf p)}\hat{\mathbf v}_{1_l}|^2.
\end{equation}

As the maximization of (86) and (87) still involves a $Q$-dimensional maximization over $\mathbf P $, further reduction in the computational load can be achieved by using the AP algorithm presented in (45)-(47), with the required modification of the cost function $g(\mathbf P)$.

\subsection { Low Complexity MUSIC-liked MVDR-like Solutions}
We next present solutions which eliminate the $Q$-dimensional serach, by resorting to MUSIC-like and MVDR-like techniques. These solutions are applicable only in  case the signals are \textit{noncoherent}

Before we present these solutions, it would be instructive to present the basis of the existing MUSIC-like and MVDR-like solutions [],[],[],[], so as to better understand the differences.

To this end,  note that from (63) we can express the signal component $\mathbf Y$ as
\begin{equation}
\mathbf Y =
\begin{pmatrix}  \mathbf A_1 \mathbf B_1\\
\vdots  \\
\mathbf A_L\mathbf B_L  \\
\end{pmatrix}\mathbf S=
\begin{pmatrix}  \tilde{\mathbf a}_1(\mathbf p_1)b_{1,1}, \dots,\tilde{\mathbf a}_1(\mathbf p_Q)b_{1,Q}  \\
\vdots\;\;\;\;\;\;\;\;\;\;\;\;\;\;\;\;\;\;\;\;\vdots  \\
\tilde{\mathbf a}_L(\mathbf p_1)b_{L,1} ,\dots,\tilde{\mathbf a}_L(\mathbf p_Q)b_{L,Q}  \\
\end{pmatrix}\mathbf S.
\label{eq:snapshot}
\end{equation}
As is evident from this structure, the signal subspace, i.e., the space  spanned by the columns of $\mathbf Y$, is spanned by $Q$ columns  having the following form:

\begin{equation}
\begin{pmatrix}  
\tilde{\mathbf a}_1(\mathbf p)b_1  \\
\vdots \\
\tilde{\mathbf a}_L(\mathbf p)b_L  \\
\end{pmatrix}.
\label{eq:snapshot}
\end{equation}
Note that this characterization of the signal subspace is parametrized by the $L$ \textit {unknown} parameters $b_1,\ldots ,b_L$, which need to be estimated from the data. This estimation step complicates the exsisting MUSIC-like and MVDR-like solutions,  presented in [],[],[],[], since it requires an eigen-decomposition of an $L\times L$ matrix for each potential location $\mathbf p$.

Alternatively, our solution is based on a different parametrization of the  signal subspace  given by (63)-(64) and based on the block-diagonal matrix $\mathbf A$, which we can  rewrite as

\begin{equation}
\mathbf A 
=\begin{pmatrix}  \tilde{\mathbf a}_1(\mathbf p_1),\ldots,\tilde{\mathbf a}_1(\mathbf p_Q) & \cdots & \mathbf 0 \\
\vdots & \ddots & \vdots \\
\mathbf 0 & \cdots  & \tilde{\mathbf a}_L(\mathbf p_1),\ldots,\tilde{\mathbf a}_L(\mathbf p_Q)\\
\end{pmatrix}
\label{eq:snapshot}
\end{equation}
As is evident,   $\mathbf A$ is parametrized only by the unknown locations. For each location $\mathbf p$, it contains a set of $L$ columns   having the following form: 
\begin{equation}
\{\overline{\mathbf a}_1(\mathbf p),\ldots,\overline{\mathbf a}_L(\mathbf p)\} ,
\end{equation}
where $\overline{\mathbf a}_1(\mathbf p)$ is a block vector with all zeros except the $l$-th block, which value is $\tilde{\mathbf a}_l(\mathbf p)$:
\begin{equation}
\overline{\mathbf a}_1(\mathbf p)
=\begin{pmatrix}  \tilde{\mathbf a}_1(\mathbf p)  \\
\mathbf 0\\
\vdots \\
\mathbf 0  \\
\end{pmatrix}
;\;\;\; \overline{\mathbf a}_L(\mathbf p)
=\begin{pmatrix}  \mathbf 0  \\
 \vdots  \\
 \mathbf 0\\
 \tilde{\mathbf a}_L (\mathbf p)\\
\end{pmatrix}.
\label{eq:snapshot}
\end{equation}

Now,  if the signals are \textit{noncoherent}, it follows from the  well-known MUSIC technique [] that  
\begin{equation}
span \mathbf A \approx span \mathbf (\hat{\mathbf v}_1,\ldots, \hat{\mathbf v}_Q),
\end{equation}
which implies that the columns of $\mathbf A$ are approximately orthogonal to the noise subspace
\begin{equation}
\mathbf A^H[(\hat{\mathbf v}_{Q+1},\ldots,\hat{\mathbf v}_M)]\approx \mathbf 0.
\end{equation}
Thus,  for each location $\mathbf p$, the set of columns given by (85) are approximately orthogonal to the noise subspace. This, in turn, implies  that the locations 
$\{\mathbf p_q\}_{q=1}^Q$ can be obtained by searching for the $Q$ highest maxima of the following function:
\begin{equation}
f(\mathbf p)= {1\over{\sum_{l=1}^L\sum_{i=Q+1}^M|\hat{\mathbf v}_i^H\overline{\mathbf a}_l(\mathbf p)|^2}}
\end{equation}
Using (93), this can be rewritten as
\begin{equation}
f(\mathbf p)= {1\over{\sum_{l=1}^L\sum_{i=Q+1}^M|\hat{\mathbf v}_{i_l}^H\tilde{\mathbf a}_l(\mathbf p)|^2}}
\end{equation}
Note that this solution is computationally much simpler than  the MUSIC-like solutions of See and Gershman [] and Weiss and Amar [] since here \textit{no eigen-decomposition is needed }  for each potential location $\mathbf p$.

Similarly, a MVDR-like  solution can be  obtained by searching  for the highest maxima of the following function:

\begin{equation}
g(\mathbf p)= {1\over{\sum_{l=1}^L\overline{\mathbf a}_l^H(\mathbf p)\hat{\mathbf R}^{-1}\overline{\mathbf a}_l(\mathbf p)}}
\end{equation}
Using (93), this can be rewritten as
\begin{equation}
g(\mathbf p)= {1\over{\sum_{l=1}^L\tilde{\mathbf a}_l^H(\mathbf p)(\hat{\mathbf R}^{-1})_{l,l}\tilde{\mathbf a}_l(\mathbf p)}}
\end{equation}
where $(\hat{\mathbf R}^{-1})_{l,l}$ denotes the $l,l$ block of  $(\hat{\mathbf R}^{-1})$.
Note that this solution is computationally much simpler than  the MVDR-like solutions of Marvychev et al [] and Tirer and Weiss [] since here  \textit{no  eigen-decomposition is needed}  for each potential location $\mathbf p$.

\section{Simulation Results}

\section{Conclusions}

\appendices  



\ifCLASSOPTIONcaptionsoff
  \newpage
\fi




\appendix
\subsection{Exact Maximum Likelihood Solution}

\indent
In this Appendix we derive the exact maximum likelihood solution.

Using the following identity [],

\begin{equation}
tr(\mathbf D^{*}_{\mathbf y} \mathbf A \mathbf D_{\mathbf z}\mathbf B^{T})=\mathbf y^{H}(\mathbf A \circ\mathbf B)\mathbf z,
\end{equation}
where  $\mathbf D_{\mathbf z}=diag(\mathbf z)$ and $\circ$ denotes the Hadamard product, we can rewrite (14), dropping the terms which do not contain $\mathbf B_l$,  as
\begin{eqnarray}
\label{eq:CM_criterion}
J_l & =  &  - \mathbf 1^H(\mathbf A_{l}\circ(\mathbf S\mathbf X^{H}_{l})^{T})\mathbf b_l-\mathbf b_l^H(\mathbf A_l^{H}\circ(\mathbf X_l\mathbf S^H))^T\mathbf 1 \nonumber\\ & + &
\mathbf b^{H}_{l}(\mathbf A^{H}_{l}\mathbf A_{l})\circ (\mathbf S\mathbf S^{H})^{T})\mathbf b_l.
\end{eqnarray}
where $\mathbf 1$ denotes the vector composed of all 1s. Taking now the derivative of $J$ with respect to $\mathbf b_l$, using  the  well known complex differentiation rule [],  and equating it to zero, we get 
\begin{equation}
\mathbf 1^{H}\mathbf A_{l}\circ(\mathbf S\mathbf X^{H}_{l})^{T} = \mathbf b^{H}_{l}(\mathbf A^{H}_{l}\mathbf A_{l})\circ(\mathbf S\mathbf S^{H})^{T},
\end{equation}
whose solution is 
\begin{equation}
\hat {\mathbf b}_l= ((\mathbf A^{H}_{l}\mathbf A_{l})\circ(\mathbf S\mathbf S^{H})\mathbf )^{-1}(\mathbf A_{l}^H\circ(\mathbf X_{l}^H\mathbf S)^T\mathbf )\mathbf 1.
\end{equation}
Now, using the following identity [],
\begin{equation}
diag( \mathbf A \mathbf D_{\mathbf z}\mathbf B^{T})=(\mathbf A \circ\mathbf B)\mathbf z,
\end{equation}
this becomes
\begin{equation}
\hat {\mathbf b}_l= ((\mathbf A^{H}_{l}\mathbf A_{l})\circ(\mathbf S\mathbf S^{H})\mathbf )^{-1}
diag(\mathbf A_l^H \mathbf X_l \mathbf S^H)
\end{equation}
Substituting this  back into (101), after some straightforward manipulation, we get
\begin{equation}
J_l=- diag(\mathbf A_l^H \mathbf X_l \mathbf S^H)^H((\mathbf A^{H}_{l}\mathbf A_{l})\circ(\mathbf S\mathbf S^{H})\mathbf )^{-1}
diag(\mathbf A_l^H \mathbf X_l \mathbf S^H)
\end{equation}
Substituting this result into (12), we get that the exact maximum likelihood solution is given by the following maximization problem
\begin{equation}
\argmax_{\mathbf S, \mathbf P}\sum_{l=1}^L diag(\mathbf A_l^H(\mathbf P) \mathbf X_l \mathbf S^H)^H \mathbf C_l^{-1}
diag(\mathbf A_l^H(\mathbf P) \mathbf X_l \mathbf S^H)
\end{equation}
where
\begin{equation}
\mathbf C_l=(\mathbf A_l^{H}(\mathbf P)\mathbf A_l(\mathbf P))\circ(\mathbf S^{*}\mathbf S^{T})\mathbf )^{-1}
\end{equation}
This solution can be interpreted as weighted correlation between $\mathbf A_l^H(\mathbf P)\mathbf X_l$ and $\mathbf S^H$, weighted by the matrix $\mathbf C_l^{-1}$. 
Unfortunately,  it  does not seem to enable further  elimination of the unknown matrix $\mathbf S$.

\subsection{An Equality Regarding the Eigenvalues }
In this Appendix we prove equation (34) regarding the equality of the eigenvalues of $\sum_{l=1}^{L} {\mathbf X_l^{H}} \mathbf{  P}_{\mathbf A_l}\mathbf X_l $ and $\mathbf P_{\mathbf A(\mathbf P)} \hat{\mathbf R} \mathbf P_{\mathbf A(\mathbf P)}$.

To this end,  note first that using the properties of the projection matrix, we have
\begin{equation}
\label{eq:W} 
 \sum_{l=1}^{L} {\mathbf X_l^{H}} \mathbf{  P}_{\mathbf A_l}\mathbf X_l =  \sum_{l=1}^{L} (\mathbf{ P}_{\mathbf A_l}\mathbf X_l)^{H} \mathbf{  P}_{\mathbf A_l}\mathbf X_l
\end{equation}
Now, since the eigenvalues of a product two matrices are unchanged by their permutation, we have
\begin{eqnarray}
 \lambda_q ( \sum_{l=1}^{L} {\mathbf D_l^{H}} \mathbf{  D}_l) 
&=& \lambda_q (\left( \begin{array}{ccc}\mathbf D_1^{H} & \cdots & \mathbf {D}_L^{H} \end{array} \right)
\left( \begin{array}{c}\mathbf D_1 \\
\vdots  \\
\mathbf D_{L} \\
\end{array}\right))\cr
&=&\lambda_q(\left( \begin{array}{c}\mathbf D_1 \\
\vdots  \\
\mathbf D_{L} \\
\end{array}\right)
\left( \begin{array}{ccc}\mathbf D_1^{H} & \cdots & \mathbf {D}_L^{H} \end{array} \right))\cr
&=&\lambda_q\begin{pmatrix}\mathbf D_1\mathbf D_1^{H} & \cdots &  \mathbf D_L \mathbf D_1^{H} \\
\vdots & \ddots & \vdots \\
\mathbf D_L\mathbf D_1^{H} & \cdots & \mathbf D_L\mathbf D_L^{H} \\
\end{pmatrix}
\end{eqnarray}
Using this identity, we have
\begin{equation}
\label{eq:W} 
\lambda_q ( \sum_{l=1}^{L} { \mathbf{  P}_{\mathbf A_l}\mathbf X_l)^{H}} \mathbf{  P}_{\mathbf A_l}\mathbf X_l)= \lambda_q (\left(\begin{array}{ccc}\tilde {\mathbf X}_1 \tilde {\mathbf X}_1^{H} & \cdots & \tilde {\mathbf X}_1\tilde {\mathbf X}_L^{H} \\
\vdots & \ddots & \vdots \\
\tilde {\mathbf X}_L \tilde {\mathbf X}_1^{H} & \cdots & \tilde {\mathbf X}_L \tilde {\mathbf X}_L^{H}  \\
\end{array}\right) )
\end{equation}
where $\tilde{ \mathbf X}_l$ is the $M_l\times N$ matrix  
\begin{equation}
\label{eq:R_i_test}
{\tilde{ \mathbf X}_l= \mathbf{  P}_{\mathbf A_l}\mathbf X_l}. 
\end{equation}
Now,  as can readily be verified,   
\begin{equation}
\label{eq:W} 
 \begin{pmatrix}\tilde {\mathbf X}_1 \tilde {\mathbf X}_1^{H} & \cdots & \tilde {\mathbf X}_1\tilde {\mathbf X}_L^{H} \\
\vdots & \ddots & \vdots \\
\tilde {\mathbf X}_L \tilde {\mathbf X}_1^{H} & \cdots & \tilde {\mathbf X}_L \tilde {\mathbf X}_L^{H}  \\
\end{pmatrix}=\mathbf P_{\mathbf A}
\mathbf X\mathbf X^H\mathbf P_{\mathbf A}=\mathbf P_{\mathbf A}\hat {\mathbf R }\mathbf P_{\mathbf A}  
\end{equation}
Combining (109)-(113), we get
\begin{equation}
\label{eq:W} 
\lambda_q(\sum_{l=1}^{L} {\mathbf X_l^{H}} \mathbf{  P}_{\mathbf A_l(\mathbf P)}\mathbf X_l )=\lambda_q ( \mathbf P_{\mathbf A(\mathbf P)} \hat{\mathbf R} \mathbf P_{\mathbf A(\mathbf P)}),
\end{equation}
which is (34).

\subsection{ Beamforming-Based Signal Subspace Solution }
\indent
In this Appendix we present a beamforming-based derivation of  the signal subspace   solution for the coherent signals case. The single signal case is a special case of coherent signals corresponding to $Q=1$

Note that when the signals are coherent, (2) can be rewritten as 

\begin{equation}
\mathbf x_{l}(t) = \mathbf A_{l}\mathbf b_{l} s(t) +\mathbf n(t),
\label{eq:snapshot}
\end{equation}

Observing this expression it is clear that the optimal weight vector for   beamforming at the $l$-th subarray, which will enable coherent summation across the subarrays,  is given by 
\begin{equation}
\mathbf w_l=\mathbf A_{l}\mathbf b_{l}.
\end{equation} 
Yet, since $\mathbf b_l$ is unknown, we need to estimate it from the  data.
To this end, first note that in the absence of noise $\mathbf x_l(t)$ spans a rank-1 subspace given  by $\mathbf A_{l}\mathbf b_{l}$. This subspace is well approximated by the largest eigenvector of $\hat{\mathbf R}_{l,l}$, denoted by $\overline{\mathbf v}_{1_l}$. Now, as we show in Appendix D

 \begin{equation}
 \overline{\mathbf v}_{1_l}=\mathbf v_{1_l}.
\end{equation} 
Thus, a natural way to estimate $\mathbf b_l$ is by    the following least squares criterion:
\begin{equation}
\label{eq:R_i_test}
{\hat {\mathbf  b}_{l}} =\argmin_{\mathbf b_{l}} |\hat{\mathbf v}_{1_l}-\mathbf A_l \mathbf b_l  |^{2},  
\end{equation}
whose solution is given by
\begin{equation}
\label{eq:R_i_test}
\hat {\mathbf b}_l = (\mathbf A_l^{H}  \mathbf A_l)^{-1}\mathbf A_l^{H}  \hat{\mathbf v}_{1_l}  
\end{equation}
Substituting this into  (116), we get
\begin{equation}
\hat {\mathbf w}_l=\mathbf A_{l}\hat {\mathbf b}_{l}=\mathbf A_{l}(\mathbf A_l^{H}  \mathbf A_l)^{-1}\mathbf A_l^{H} \mathbf {\hat{\mathbf v}}_{1_l}=\mathbf P_{\mathbf A_l}\hat{\mathbf v}_{1_l}.
\end{equation}

Using this vector for bramforming at the $l$-th subarray, the location $\mathbf p$ for which the  sum of the beamformers' power output  is maximized is given by
\begin{equation}
\label{eq:W} 
\mathbf{\hat  P} =\argmax_{\mathbf P} \sum_{n=1}^{N}|\sum_{l=1}^{L} ( \mathbf{  P}_{\mathbf A_l(\mathbf P)}\hat{\mathbf v}_{1_l})^{H} \mathbf x_{l}(t_n)|^{2} 
\end{equation}
which is  (81).  

\subsection{ The Case of Rank-1 Covariance Matrix}

In this Appendix we prove the relation (112) between the array and the subarrays covariance matrices in case the array covariance matrix is rank-1.

To this end, note that if the array covariance matrix is rank-1 then we have
\begin{equation}
\mathbf R -\sigma^2 \mathbf I_M =\mathbf v_1\mathbf v_1^{H}=\begin{pmatrix}\mathbf v_{1_1} \mathbf v^{H}_{1_1} & \cdots & \mathbf v_{1_1}^H\mathbf v^{H}_{1_L}  \\
\vdots & \ddots & \vdots \\
\mathbf v_{1_L}\mathbf v_{1_1}& \cdots  & \mathbf v_{1_L}^H\mathbf v^{H}_{1_L}  \\
\end{pmatrix},
\label{eq:snapshot}
\end{equation}
where $\mathbf v_{1_l}$ is the $l$-th subvector of $\mathbf v_1$.
Now, for the rank-1 case we also have
\begin{equation}
\mathbf R_{l,l} -\sigma^2 \mathbf I_{M_l}=\overline{\mathbf v}_{1_l}\overline{\mathbf v}_{1_l}^H,
\label{eq:snapshot}
\end{equation}
Since  the $l,l$ element of $\mathbf R-\sigma^2 \mathbf I_M$ should be equal to $\mathbf R_{l,l}-\sigma^2\mathbf I_{M_l}$, it follows that
 \begin{equation}
\overline{ \mathbf v}_{1_l}=\mathbf v_{1_l}.
\end{equation} 
which is (117).

\end{document}